\documentclass[aps,pra,10pt,reprint,superscriptaddress]{revtex4-2}

\usepackage{natbib}
\usepackage{amsmath,amsfonts,amssymb}
\usepackage{xcolor,graphicx}
\usepackage{tikz}
\usepackage{braket}
\usepackage{dcolumn}
\usepackage{bm}
\usepackage{subfig}
\usepackage{float}
\usepackage[pdftex]{hyperref}
\usepackage{csquotes}
\hypersetup{
	colorlinks = true,
	linkcolor = blue,
	anchorcolor = blue,
	citecolor = red,
	filecolor = blue,
	urlcolor = blue}

\DeclareMathOperator{\ii}{\textrm{i}} 
\begin{document}

\title{Effects of classical drivings on  the power broadening of atomic lineshapes}

\author{L. Hernández-Sánchez}
\email[e-mail: ]{leonardi1469@gmail.com}
\affiliation{Facultad de Ciencias en F\'isica y Matem\'aticas, Universidad Aut\'onoma de Chiapas, Carretera Emiliano Zapata, Km. 8, Rancho San Francisco, 29050 Tuxtla Guti\'errez, Chiapas, Mexico}
\affiliation{Centro de Bachillerato Tecnológico Industrial y de Servicios No. 144: José Emilio Grajales Moguel, Boulevard Mayor Sabines No. 1982, Col. 24 de Junio, Tuxtla Guti\'errez, Chiapas, 29047, Mexico}
\author{I. A. Bocanegra-Garay}
\affiliation{Departamento de F\'isica Te\'orica, At\'omica y \'Optica and  Laboratory for Disruptive Interdisciplinary Science, Universidad de Valladolid, 47011 Valladolid, Spain}
\author{I. Ramos-Prieto}
\affiliation{Instituto Nacional de Astrofísica Óptica y Electrónica, Calle Luis Enrique Erro No. 1\\ Santa María Tonantzintla, Puebla, 72840, Mexico}
\author{F. Soto-Eguibar}
\affiliation{Instituto Nacional de Astrofísica Óptica y Electrónica, Calle Luis Enrique Erro No. 1\\ Santa María Tonantzintla, Puebla, 72840, Mexico}
\author{H. M. Moya-Cessa}
\affiliation{Instituto Nacional de Astrofísica Óptica y Electrónica, Calle Luis Enrique Erro No. 1\\ Santa María Tonantzintla, Puebla, 72840, Mexico}

\date{\today}

\begin{abstract}
In the framework of the Jaynes-Cummings model, we investigate how  atomic lineshapes are affected by coherently driving the atom-field interaction. We pay particular attention to the two-level atom interaction with a thermal cavity field,  when both are influenced by  external classical fields. Adopting a density matrix formalism, we calculate the average atomic inversion and demonstrate how the corresponding lineshapes vary as a function of the average number of thermal photons, and the atom-field classical coupling. Furthermore, we compare these results with those obtained from the standard Jaynes-Cummings model and validate our findings through numerical calculations.
\end{abstract}
\maketitle

\section{Introduction}
Over the past few decades, the Jaynes-Cummings model (JCM) has emerged as a fundamental cornerstone in the development of quantum optics~\cite{JC_1963}. This model describes the interaction between electromagnetic radiation and matter in its simplest form, considering only the relationship between a single mode of radiation and a two-level atom, which enables a simple and elegant mathematical formulation. Due to this simplicity, the model has been extensively studied from both theoretical and experimental perspectives, providing a solid foundation for understanding fundamental quantum-optical phenomena over the years~\cite{JC_1963,Meschede_1985,Filipowicz_1986,Bruce_1993}.

However, the field of quantum optics has encountered more involved and realistic scenarios that demand additional considerations. In consequence, generalizations of the standard JCM have been developed~\cite{Gerry_2004,Walls_2008,Klimov_2009}; these generalizations span various areas, including extensions to multiple atomic levels and field modes~\cite{Sukumar_1981,Kochetov_1988,Buzek_1990,Abdalla_1991}, as well as the incorporation of nonlinear effects~\cite{Abdala_1990,Rodriguez_2013,Santos_2012,Santos_2016}, among others~\cite{Phoenix_1991,Moya_1995a,Ramos_2014,Ramos_2020,Medina_2020}. Furthermore, a wide variety of initial conditions have been explored for both the atomic and field subsystems, encompassing semi-classical and non-classical states, such as: coherent states~\cite{Bruce_1993,Eberly_1980,Gerry_2004}, superpositions of coherent states~\cite{Buzek_1992,Buzek_1992b,Moya_1995b,Segundo_2003,Muhammad_2009,Hernandez_2023_a}, squeezed coherent states~\cite{Moya_1992,Alebachew_2007}, as well as diverse combinations of these~\cite{Vidiella_1995,Hernandez_2023_b}.

The standard JCM predicts the emergence of entanglement between the two-level system and the cavity field~\cite{Alsing_1992}. Previous research has demonstrated that an appropriate way to analyze this phenomenon is by means of an additional external classical field~\cite{Alsing_1992,Dutra_1994,Bocanegra_2023}. This approach led Alsing et al.~\cite{Alsing_1992} to conduct a theoretical study of a driven JCM, which was later experimentally verified by Thompson and his team~\cite{Thompson_1992}. In the new model an external classical electromagnetic field driving the system is considered. Adding this external field has a significant impact on the dynamics of the system, resulting in a plethora of new quantum phenomena~\cite{Alsing_1992,Dutra_1994,Bocanegra_2023}.

On the other hand, spectral lineshapes are pivotal in spectroscopy, furnishing vital insights into the interplay between matter and electromagnetic radiation~\cite{Svanberg_1992}. However, these lineshapes are susceptible to various influences, such as nearby non-resonant levels~\cite{Moya_1991,Hernandez_2023_a,Hernandez_2023_b}, as evidenced in experimental studies with micromasers~\cite{Filipowicz_1986}. Moreover, recent research has underscored the significant impact of atom temperature and trapping effects on these spectral characteristics~\cite{Shih_2013, Bianchet_2022}. Additionally, temperature also plays a critical role; at lower temperatures, thermal asymmetries become less discernible~\cite{Rempe_1990a,Rempe_1990b}, emphasizing the influence of environmental factors on these spectral features.

In this study, we focus on analyzing a driven JCM, which allows for the simultaneous excitation of both the atom and the cavity field mode due to the presence of an external classical field~\cite{Bocanegra_2023, Hernandez_book}. Our main objective is to investigate how the external classical field influences the shapes of atomic lines when the initial state of the quantum field is thermal and the atom is initially in the excited state. Through this research, we aim to highlight the importance of atomic lineshapes in the field of quantum optics and spectroscopy, and their utility as powerful tools for exploring complex quantum systems. Additionally, we aim to provide exact analytical results detailing the effect that an external classical driving field produces on them.

\section{The driven Jaynes-Cummings Model}\label{Sec_A}
\begin{figure}[H]
\centering
\includegraphics[width=.85\linewidth]{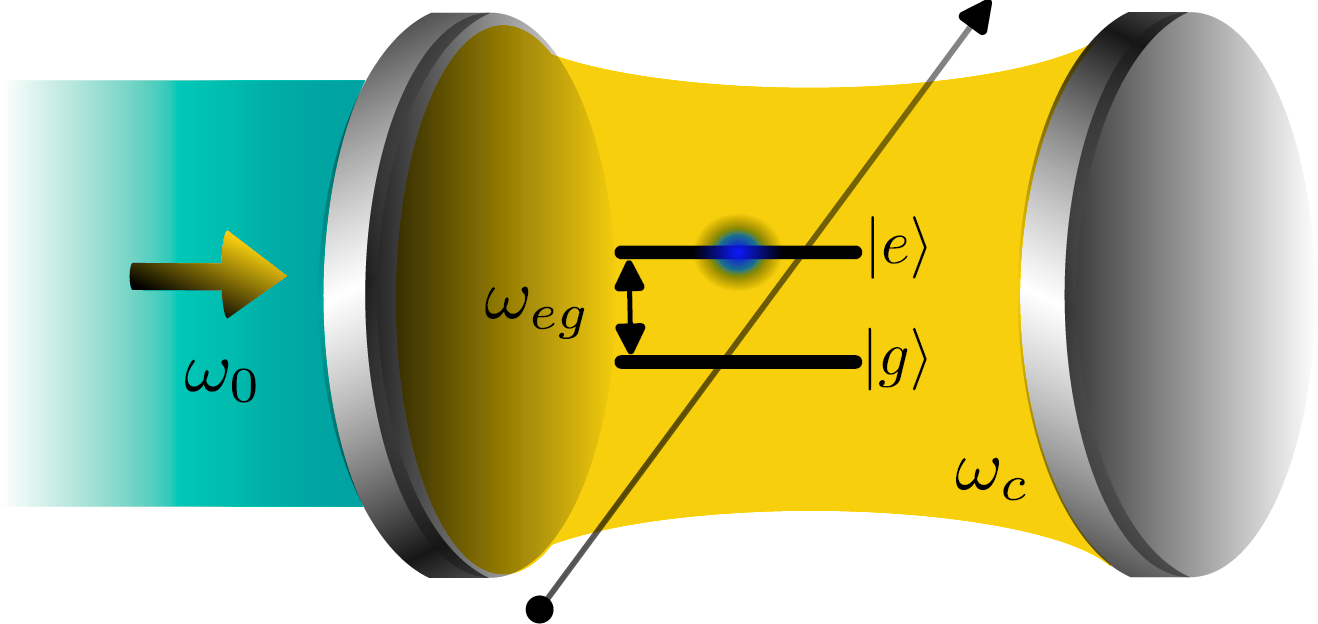}
\caption{Scheme of a lossless cavity formed by perfectly reflecting mirrors. Within the space between the mirrors, a two-level atom with a transition frequency $\omega_{eg}$ interacts with a cavity field of frequency $\omega_c$. Additionally, both the atom and the cavity field, are influenced by an external classical field with frequency $\omega_0$.}
\label{fig1}
\end{figure}

Let us consider a system consisting of a two-level atom, with states denoted as $\ket{g}$ (ground state) and $\ket{e}$ (excited state), having a transition frequency $\omega_{eg}$. The atom is placed within a cavity formed by perfectly reflecting mirrors sustaining a single quantized electromagnetic field mode with frequency $\omega_{c}$. Additionally, we consider an external classical field with frequency $\omega_{0}$, driving both the atom and the cavity field. This setup is depicted in Fig.~\ref{fig1}.
The time-dependent Hamiltonian that describes this system can be written as~\cite{Alsing_1992,Bocanegra_2023} 
\begin{equation}\label{H_0}
 \begin{split}
\hat{\mathcal{H}}  & = \; \frac{\omega_{eg}}{2} \hat{\sigma}_{z} +  \omega_c\hat{a}^{\dagger} \hat{a} + g \left( \hat{\sigma}_{+}\hat{a} + \hat{\sigma}_{-} \hat{a}^{\dagger} \right) \\ &  +  \zeta\left( \hat{\sigma}_-e^{\ii\omega_0t} + \hat{\sigma}_+e^{-\ii\omega_0t} \right) + \xi \left( \hat{a}e^{\ii\omega_0t}+\hat{a}^{\dagger} e^{-\ii\omega_0t} \right),
 \end{split}
\end{equation}
where the real parameters $g$, $\zeta$ and $\xi$ are the coupling constants between the  atom and the cavity field, the external classical field and the atom, and the cavity and classical fields, respectively. Furthermore, throughout this work, we adopt the convention that $\hbar = 1$. As is customary, we consider the creation and annihilation operators $\hat{a}^{\dagger}$ and $\hat{a}$, which satisfy the usual bosonic commutation relation $[\hat{a}, \hat{a}^{\dagger}] = 1$, to describe the cavity field mode.  In turn, for the atomic part of the system, we employ the operators $\hat{\sigma}_{+} = \ket{e}\bra{g}$, $\hat{\sigma}_{-} = \ket{g}\bra{e}$, and $\hat{\sigma}_{z} = \ket{e}\bra{e} - \ket{g}\bra{g}$, which satisfy the commutation relations $[\hat{\sigma}_{+}, \hat{\sigma}_{-}] = \hat{\sigma}_{z}$ and $[\hat{\sigma}_{z} , \hat{\sigma}_{\pm}] = \pm 2 \hat{\sigma}_{\pm}$.

The dynamics of the system associated to the Hamiltonian in~\eqref{H_0} is governed by the Schrödinger equation
\begin{equation}\label{Schrodinger_0}
    \ii \frac{\partial \ket{\psi(t)}}{\partial t}=\hat{\mathcal{H}}\ket{\psi(t)}.
\end{equation}
Through an invariant approach, it has been demonstrated in reference~\cite{Bocanegra_2023,Hernandez_book} that employing the pair of unitary transformations $\hat{T}=\exp\left[\ii \omega_{0} t (\hat{n}+ \hat{\sigma}_z /2)\right]$ and $\hat{D}(\alpha) = \exp\left[\alpha (\hat{a}^\dagger - \hat{a})\right]$, where $\hat{n} = \hat{a}^\dagger\hat{a}$ is the photon-number operator and $\hat{D}(\alpha)$ is the displacement operator~\cite{Gerry_2004,Bruce_1993, Walls_2008} with $\alpha = \zeta / g$  (see also Ref. \cite{Bocanegra_2024} for the implementation of displacement operators in a different context), it is possible to transform the Hamiltonian of the driven system into the Hamiltonian of the standard JCM, namely
\begin{equation}\label{H_JCM}
\hat{\mathcal{H}}_{\mathrm{JC}} = \Delta_{c}\hat{n}+\frac{\Delta_{eg}}{2} \hat{\sigma}_z + g\left(\hat{\sigma}_{+} \hat{a}  + \hat{\sigma}_{-} \hat{a}^{\dagger} \right),
\end{equation}
where $\Delta_{c} = g\xi/\zeta = \omega_c - \omega_0$ and $\Delta_{eg} = \omega_{eg} - \omega_0$ represent the detunings between the cavity and classical fields, and the atomic and classical fields, respectively. This straightforwardly leads to the general solution of the Schr\"odinger equation~\eqref{Schrodinger_0} as follows
\begin{equation}\label{psi}
\ket{\psi (t)} =  \hat{T}^{\dagger} \hat{D}^{\dagger} (\alpha) \hat{U}_{\mathrm{JC}}(t) \hat{D} (\alpha) \ket{\psi (0)},
\end{equation}
where $\hat{U}_{\mathrm{JC}}(t)$ is the evolution operator associated with the standard Jaynes-Cummings Hamiltonian, given by~\cite{Buzek_1992,Gerry_2004,Klimov_2009}
\begin{equation}\label{U_JCM}
\hat{U}_{\mathrm{JC}} (t) = e^{-\ii t \hat{\mathcal{H}}_{\mathrm{JC}}} 
= e^{-\ii \Delta_{c} t (\hat{n}+ \hat{\sigma}_z /2)}
\begin{bmatrix} 
\hat{U}_{11}(t) & \hat{U}_{12}(t) 
\\ 
\hat{U}_{21}(t)  & \hat{U}_{22}(t)  
\end{bmatrix},
\end{equation}
where
\begin{subequations}\label{Op_Evol}
\begin{align}
    \label{OE_1}
     \hat{U}_{11} (t) & = \cos \left( \Omega_{\hat{n}+1} t \right) - \ii \frac{\Delta}{2} \frac{\sin \left( \Omega_{\hat{n}+1} t \right)}{\Omega_{\hat{n}+1}},\\
    \label{OE_2}
     \hat{U}_{12} (t) & =  - \ii g \hat{a} \frac{\sin \left( \Omega_{\hat{n}} t \right)}{\Omega_{\hat{n}}}, \\
    \label{OE_3}
     \hat{U}_{21} (t) & =  - \ii g \hat{a}^{\dagger} \frac{\sin \left( \Omega_{\hat{n}+1} t \right)}{\Omega_{\hat{n}+1}}, \\
    \label{OE_4}
     \hat{U}_{22} (t) & = \cos \left(\Omega_{\hat{n}} t \right) + \ii \frac{\Delta}{2} \frac{\sin \left( \Omega_{\hat{n}} t \right)}{\Omega_{\hat{n}}},
\end{align}
\end{subequations}
and
\begin{equation}\label{F_Rabi}
    \Omega_{\hat{n}} =  \sqrt{\frac{\Delta^2}{4} + g^2 \hat{n}}, \qquad \Delta = \Delta_{eg} - \Delta_c = \omega_{eg} - \omega_{c}.
\end{equation}
It is important to note the constriction $\Delta_c = g \xi/\zeta$, meaning that there are only five free parameters out of the initial six parameters in the Hamiltonian of~\eqref{H_0}. From here on, we set $\omega_0 = \omega_c - g \xi/\zeta$.\\

Once the solution \eqref{psi} to the Schr\"odinger equation of the total system  has been obtained, it is possible to write the corresponding density matrix $\hat{\rho}(t)$ as
\begin{equation} \label{rho}
\begin{split}
\hat{\rho}(t) & = \ket{\psi(t)} \bra{\psi(t)} \\
& = \hat{T}^{\dagger} \hat{D}^{\dagger} (\alpha) \hat{U}_{\mathrm{JC}}(t) \hat{D} (\alpha) \hat{\rho}(0) 
\hat{D}^\dagger (\alpha) \hat{U}_{\mathrm{JC}}^\dagger(t) \hat{D} (\alpha) \hat{T},
\end{split}
\end{equation}
where $\hat{\rho}(0) = \hat{\rho}^F(0) \otimes \hat{\rho}^A(0)$. Here, $\hat{\rho}^F(0)$ represents the initial density matrix for the field and $\hat{\rho}^A(0)$ represents the initial density matrix for the atom, respectively.\\

The versatility of the density matrix makes it a fundamental tool for studying systems, both pure and mixed, with particular relevance in the analysis of composite or entangled systems. Its application allows the calculation of the dynamic variables of the system, requiring only the establishment of initial conditions, for both the atom and the field. From now on we focus on the scenario where the initial field is in a thermal state, a choice supported by the usual presence of a small number of photons in the cavity due to temperature in micromaser experiments~\cite{Meschede_1985}. Additionally, we consider that the atom starts in the excited state $\ket{e}$. With these premises, we calculate the atomic inversion and proceed to determine the expression for the time-averaged atomic inversion, commonly known in the literature as {\it lineshapes}. This approach will allow us to understand how temperature affects the characteristics of atomic lines, when an external classical field drives both the atom and the cavity field.

\section{Atomic inversion}\label{Atomic inversion}
From the density matrix in~\eqref{rho}, we can compute the expected value of any operator $\hat{A}$ or observable of the system using the following expression
\begin{equation} \label{V_E}
\braket{\hat{A}} = \text{Tr} \bigl[\hat{\rho} \hat{A}\bigr] = \text{Tr}\bigl[ \hat{A} \hat{\rho} \bigr].
\end{equation}

In particular, in this work, we are interested in examining the atomic inversion, defined as the difference between the probability of the atom being in the excited state and the probability of it being in the ground state. This can be calculated as the expected value of the operator $\hat{\sigma}_z$, namely
\begin{equation} \label{VE_Sigma_z1}
\begin{split}
\braket{\hat{\sigma}_z} & = \text{Tr} \left[ \hat{T}^{\dagger} \hat{D}^{\dagger} (\alpha) \hat{U}_{\mathrm{JC}}(t) \hat{D} (\alpha) \hat{\rho}(0) \hat{D}^\dagger (\alpha) \hat{U}_{\mathrm{JC}}^\dagger(t) \hat{D} (\alpha) \hat{T} \; \hat{\sigma}_z \right] \\
& = \text{Tr} \left[ \hat{U}_{\mathrm{JC}}^\dagger(t) \; \hat{\sigma}_z \;  \hat{U}_{\mathrm{JC}}(t) \hat{D} (\alpha) \hat{\rho}(0) \hat{D}^\dagger (\alpha) \right],
\end{split}
\end{equation}
where in the last equality we have employed trace properties, as well as the fact that the operators $\hat{D} (\alpha)$ and $\hat{T}$ commute with the operator $\hat{\sigma}_z$.

Assuming that initially the atom is in an excited state and the field is in a thermal state, the density matrix at $t=0$ is given by
\begin{equation}\label{Initial condition}
    \hat{\rho} (0) = \frac{1}{1+\bar{n}} \sum_{k=0}^\infty \left(  \frac{\bar{n}}{1+\bar{n}} \right)^k \ket{k,e}\bra{k,e}.
\end{equation}
Using the matrix representation of the operators and states involved, and meticulously performing the calculations, we obtain that
\begin{equation} \label{VE_Sigma_z2}
\begin{split}
\braket{\hat{\sigma}_z} & = \sum_{m=0}^\infty \bra{m} \left\lbrace \left( \hat{U}_{11}^\dagger \hat{U}_{11} - \hat{U}_{21}^\dagger \hat{U}_{21} \right) \right. \\
& \times   
\hat{D} (\alpha) \left[   \frac{1}{1+\bar{n}} \sum_{k=0}^\infty \left(  \frac{\bar{n}}{1+\bar{n}} \right)^k \ket{k}\bra{k} \right] \left. \hat{D}^\dagger (\alpha)  \right\rbrace  \ket{m} \\
&=  \frac{e^{-|\alpha|^2}}{1+\bar{n}} \sum_{m=0}^{\infty}\frac{1}{\Omega_{m+1}^2} \left[\frac{\Delta^2}{4}+g^2 \left(m+1\right) \cos\left( 2\Omega_{m+1} t\right)   \right] \\
& \times   \left\lbrace
\sum_{k=0}^{m} \left( \frac{\bar{n}}{1+\bar{n}}\right)^k
\frac{k!}{m!} \left|\alpha\right|^{2\left( m-k\right) }
\left[ L_k^{\left(m-k\right)} \left(\left| \alpha \right|^2\right)\right]^2 
\right. 
\\  & + \left. 
\sum_{k=m+1}^{\infty} \left( \frac{\bar{n}}{1+\bar{n}}\right)^k 
\frac{m!}{k!} \left|\alpha\right|^{2\left(k-m\right) }
\left[ L_m^{\left(k-m\right)} \left(\left| \alpha \right|^2\right)\right]^2  
\right\rbrace, 
\end{split}
\end{equation}
where $\Omega_{m+1} = \sqrt{\frac{\Delta^2}{4} + g^2 (m+1)}$ is the corresponding (modified) Rabi frequency and $L_\nu^{\left(\mu\right)}(z)$ are the associated Laguerre polynomials \cite{Arfken_2005}.

It is important to note that when we omit the external classical field ($\alpha = 0$), the atomic inversion reduces to the more compact expression
\begin{equation}
\braket{\hat{\sigma}_z}  = \sum_{m=0}^{\infty} \frac{\bar{n}^m}{(1+\bar{n})^{m+1}} \frac{1}{\Omega_{m+1}^2} \left[\frac{\Delta^2}{4}+g^2 \left(m+1\right) \cos\left( 2\Omega_{m+1} t\right)   \right],
\end{equation}
which corresponds to the standard JCM when the initial quantum field is a thermal field~\cite{Eberly_1980,Klimov_2009, Hernandez_book}.

\begin{figure}[H]
\centering
\includegraphics{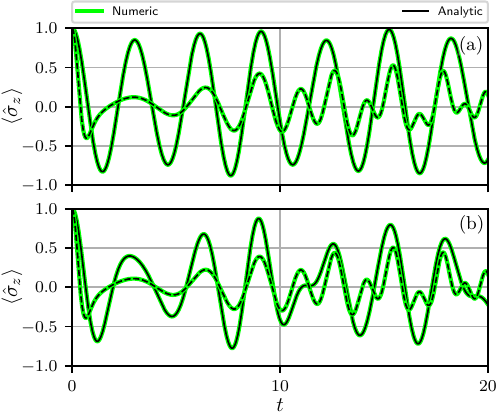}
\caption{Atomic inversion $\braket{\hat{\sigma}_z}$ corresponding to the initial condition of the atom in the excited state and the cavity field in a thermal state, using the following parameter values: $\omega_c=0.4$, $\omega_{eg}=0.9$, $g=1.0$, $\zeta=0.7$, $\xi=0.2$, $\omega_0=\omega_c-g \xi/\zeta$, with different values of the average number of photons: $\bar{n} = 0.1 $ (solid lines) and $\bar{n} = 4.0 $ (dashed lines). In (a) the atomic inversion corresponding to the standard JCM is shown. In (b) the atomic inversion corresponding to the driven JCM is displayed. The black lines represent the analytical result, while the green ones stand for the numerical solution obtained using QuTiP~\cite{Qutip}.}
\label{Inversion atomica}
\end{figure}

In Fig.~\ref{Inversion atomica}, the atomic inversion is shown for different values of the average photon number. For $\bar{n}=0.1$ (solid lines) and for $\bar{n}=4.0$ (dashed lines), along with parameter values such as $\omega_c=0.4$, $\omega_{eg}=0.9$, $g=1.0$, $\zeta=0.7$, $\xi=0.2$, and $\omega_0=\omega_c-g \xi/\zeta$. In subfigure \ref{Inversion atomica} (a), the atomic inversion for the standard JCM is depicted, while in subfigure \ref{Inversion atomica} (b), the atomic inversion for the driven JCM is displayed. From these figures, it is noticeable that in the driven JCM, the amplitude of oscillations in the atomic inversion is less pronounced than in the case of the standard JCM, indicating the effect of the external classical field on atomic transitions. Additionally, it can be observed that as the average photon number $\bar{n}$ increases, the atomic inversion for both models tends to have the same shape, suggesting that for a large average number of photons (at high temperatures), the action of the classical field on atomic transitions becomes insignificant.

Indeed, the values assigned to parameters such as the coupling strength ($g$), cavity frequency ($\omega_c$), and transition frequency ($\omega_{eg}$) directly influence the outcomes of our simulations. The parameter values are carefully calibrated to balance computational tractability with fidelity to the physical system under investigation. By offering insights into the interplay between theoretical modeling and experimental factibility, we aim to enhance the rigor of our analysis.

\section{Average atomic inversion}\label{Average atomic inversion}
In this section, we analyze how the average (in time) transition probability between the ground state and the excited state is modified as a function of the detuning. In addition, we show how this depends on both the (average) temperature and the intensity of the driving classical field. To carry out this analysis, we employ the average atomic inversion $\overline{W}(\Delta)$~\cite{Moya_1991, Hernandez_book},
\begin{equation} \label{W_Delta}
\overline{W}(\Delta) = \lim_{T \to{\infty}} \frac{1}{T} \int_{0}^{T} \braket{ \hat{\sigma}_z} dt.
\end{equation}
Since
\begin{equation}
\lim_{T \to{\infty}} \frac{1}{T} \int_{0}^{T} \cos (2\Omega_{n+1} t) \; dt=0,
\end{equation}
using \eqref{VE_Sigma_z2}, we have
\begin{equation} \label{Lineshapes}
\begin{split}
\overline{W}(\Delta)  = & \frac{e^{-|\alpha|^2}}{1+\bar{n}} \sum_{m=0}^{\infty} \left(\frac{\Delta}{2 \Omega_{m+1}} \right)^2 \\
& \times   \left\lbrace
\sum_{k=0}^{m} \left( \frac{\bar{n}}{1+\bar{n}}\right)^k
\frac{k!}{m!} \left|\alpha\right|^{2\left( m-k\right) }
\left[ L_k^{\left(m-k\right)} \left(\left| \alpha \right|^2\right)\right]^2 
\right. 
\\  & + \left. 
\sum_{k=m+1}^{\infty} \left( \frac{\bar{n}}{1+\bar{n}}\right)^k 
\frac{m!}{k!} \left|\alpha\right|^{2\left(k-m\right) }
\left[ L_m^{\left(k-m\right)} \left(\left| \alpha \right|^2\right)\right]^2  
\right\rbrace.
\end{split}
\end{equation}
In the absence of an external classical field, the average atomic inversion reduces to
\begin{equation}
\overline{W}(\Delta)  = \sum_{m=0}^{\infty} \frac{\bar{n}^m}{(1+\bar{n})^{m+1}}\left(\frac{\Delta}{2 \Omega_{m+1}} \right)^2.
\end{equation}

\begin{figure}[H]
\centering
\includegraphics{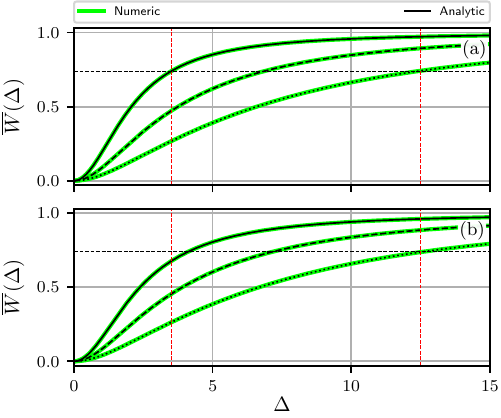}
\caption{Average atomic inversion $\overline{W}(\Delta)$ corresponding to the initial condition of the atom in the excited state and the cavity field in a thermal state, with  $g=1.0$, $\zeta=0.7$ and different values of the average photon number: $\bar{n}=0.1$ (solid lines), $\bar{n}=4.0$ (dashed lines), and $\bar{n}=15.0$ (dotted lines). In (a), the average atomic inversion corresponding to the standard JCM is shown. In (b), the average atomic inversion corresponding to the driven JCM is displayed. The black lines represent the analytical result, while the green ones stand for the numerical solution.}
\label{Inversion atomica promedio}
\end{figure}

\begin{figure}[H]
\centering
\includegraphics[width=0.45\textwidth]{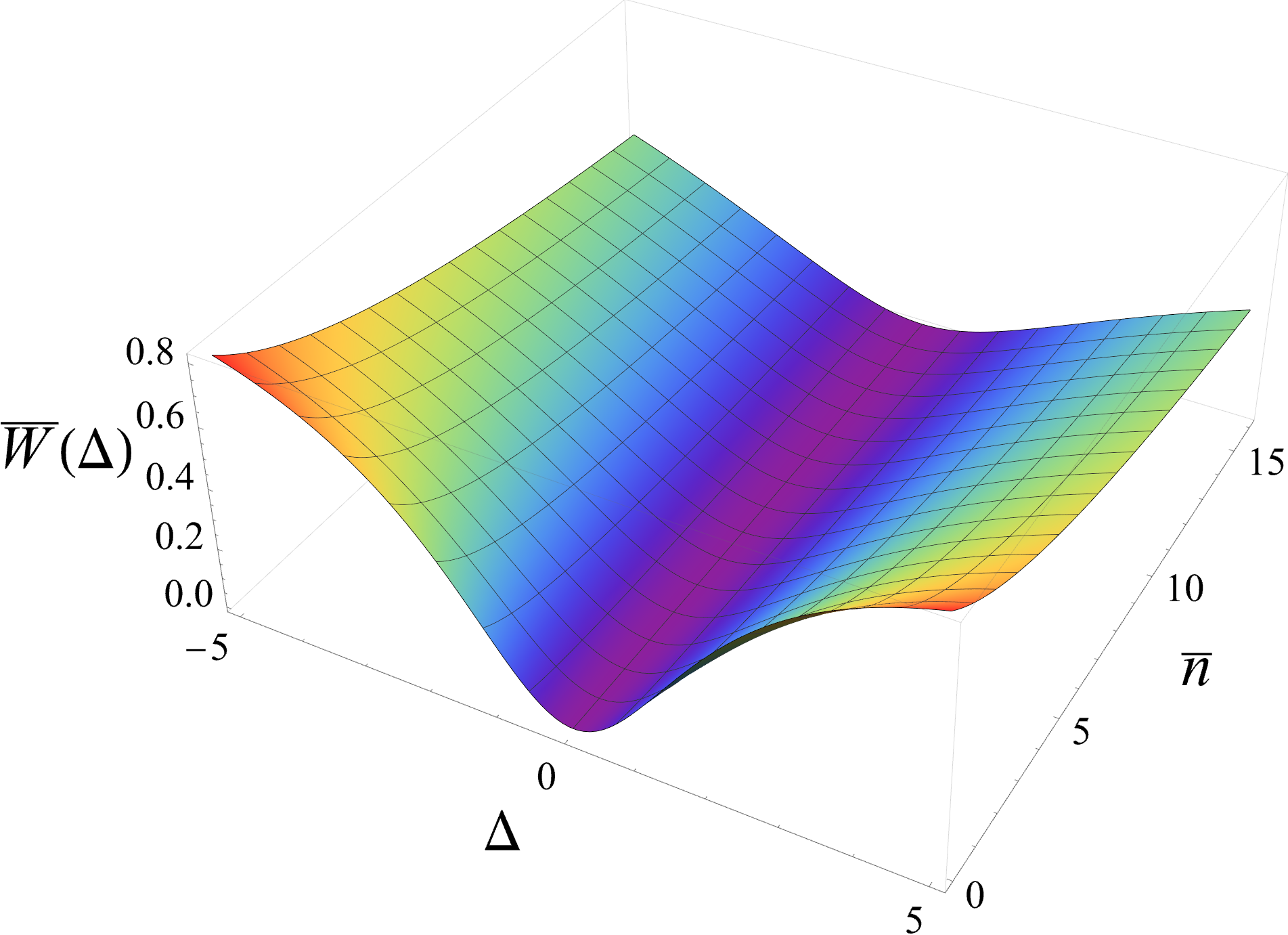}
\caption{Average atomic inversion $\overline{W}(\Delta)$ corresponding to the driven JCM as function of the average number of photons $\bar{n}$, with $\zeta=0.7$ and $g=1.0$.}
\label{Inversion atomica promedio 3d_nbar}
\end{figure}

\begin{figure}[H]
\centering
\includegraphics[width=0.45\textwidth]{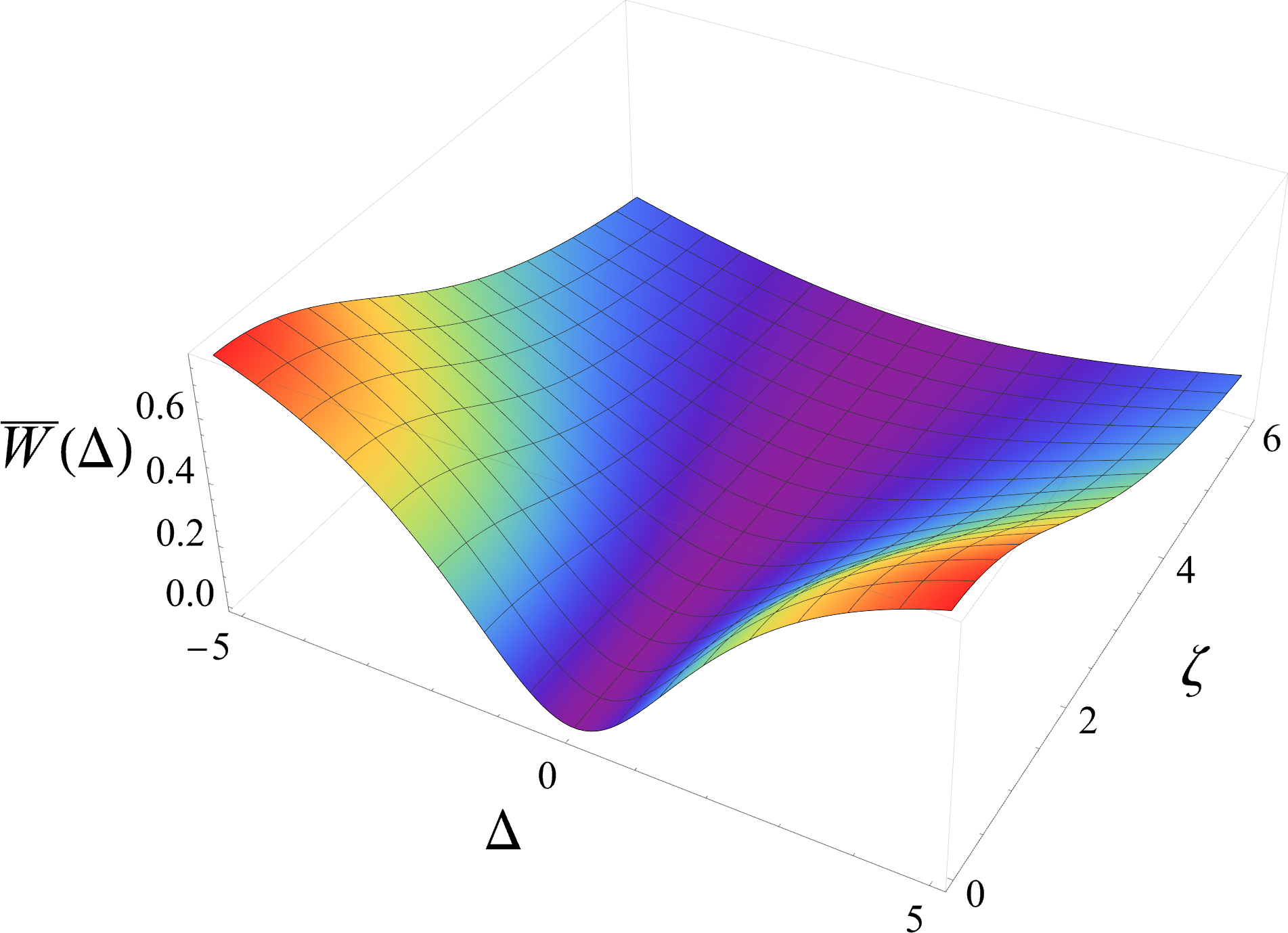}
\caption{Average atomic inversion $\overline{W}(\Delta)$ corresponding to the driven JCM as function of the parameter $\zeta$, with $\bar{n}=1.0$ and $g=1.0$. When $\zeta=0.0$, the standard JCM is recovered.}
\label{Inversion atomica promedio 3d_alpha}
\end{figure}

In Fig.~\ref{Inversion atomica promedio}, we present the average atomic inversion profiles for different values of the average photon number: $\bar{n}=0.1$ (solid lines), $\bar{n}=4.0$ (dashed lines), and $\bar{n}=15.0$ (dotted lines). These profiles are depicted with constant parameters $g=1.0$ and $\zeta=0.7$, spanning the $\Delta$ range from 0 to 15. Subfigure~\ref{Inversion atomica promedio} (a) displays the average atomic inversion for the standard JCM, while subfigure~\ref{Inversion atomica promedio} (b) illustrates the average atomic inversion for the driven JCM. We observe that as the average photon number $\bar{n}$ increases, the lineshapes broaden, with this effect being more pronounced in the driven JCM than in the standard JCM for small values of $\bar{n}$. Additionally, for relatively large values of $\bar{n}$, the lineshapes exhibit similar behavior to the driven JCM, as indicated by the dashed red lines. It is also worth noting that the lineshapes are symmetric about $\Delta =0$, as depicted in Figures \ref{Inversion atomica promedio 3d_nbar} and \ref{Inversion atomica promedio 3d_alpha}. The impact on the lines is not substantial due to the relatively small coherent driving utilized in Fig.~\ref{Inversion atomica promedio}, and this response can be finely adjusted by optimizing the parameter $\zeta$. Moreover, the broadening effect of the atomic lines for a continuous variation of the thermal average photon number $\bar{n}$ in the driven JCM can be observed in Fig.~\ref{Inversion atomica promedio 3d_nbar} for a fixed value of $\zeta$.

It is crucial to highlight that the difference between the lineshapes of the standard JCM and the driven JCM becomes more pronounced as the parameter $\alpha$ increases. Such effect is obtained, for instance, when either the amplitude of the coupling between the classical field and the atom ($\zeta$) grows or when the coupling between the cavity field and the two-level system ($g$) diminishes. This in turn can be observed in Fig.~\ref{Inversion atomica promedio 3d_alpha}, where $g$ is fixed at 1.0, the average number of photons is $\bar{n}=1.0$, and the parameter $\zeta$ is varied from 0.0 to 6.0. This results to have a pretty interesting physical consequence: as the coupling between the atom and the classical field increases, a significantly larger detuning $\Delta$ is required between the atom and the cavity field for the atom to return to its initial state. Consequently, this demonstrates how the classical driving field influences the transitions between the two levels of the atom. Furthermore, such transitions can be finely manipulated by means of the classical driving field, this last then acting as our control signal.

To conclude this manuscript, it is crucial to consider what would happen if the cavity were prepared in a different initial state. To address this question, we present an analysis of the case where the cavity initial state is configured in a Fock state $\ket{k}$, implying a well-defined number of photons in the cavity. In this context, the average atomic inversion is determined by the expression
\begin{equation}
\overline{W}(\Delta) = e^{-|\alpha|^2} \sum_{m=0}^{\infty} \left(\frac{\Delta}{2\Omega_{m+1}} \right)^2
\frac{k!}{m!} \left|\alpha\right|^{2\left( m-k\right) }
\left[ L_k^{\left(m-k\right)} \left(\left| \alpha \right|^2\right)\right]^2.
\end{equation}

\begin{figure}[h]
\centering
\includegraphics{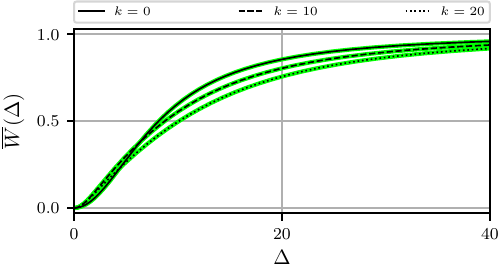}
\caption{Average atomic inversion $\overline{W}(\Delta)$ corresponding to the initial condition of the atom in the excited state and the cavity field in a Fock state, with  $g=1.0$, $\zeta=0.7$ and different values of the photon number: $k=0$ (solid lines), $k=10$ (dashed lines), and $k=20$ (dotted lines).  The black lines represent the analytical result, while the green ones stand for the numerical solution.}
\label{Fock lineshapes}
\end{figure}

In Fig.~\ref{Fock lineshapes}, we present three different cases for varying values of $k$, using the same parameter values as in the previous figures. Upon analyzing these cases, we observe that the increase in the value of $k$ leads to a progressive broadening of the atomic lines. This phenomenon suggests that, as the number of photons in the cavity increases, a greater detuning between the frequencies of the cavity and the atom is required for the latter to return to its initial state, similar to what is observed when increasing the average number of photons in the thermal state.\\
It is noteworthy that the analysis of other initial states of the cavity, such as coherent states or squeezed coherent states, is detailed in reference~\cite{Hernandez_book}.

\section{Conclusions}\label{Conclusion}

We have explored the effects of an external classical driving field on the atomic lineshapes (specifically, the average atomic inversion) within the framework of the conventional Jaynes-Cummings model. Departing from the density matrix corresponding to the exact solution of the Schr\"odinger equation for the Hamiltonian describing the driven system, we considered a specific initial condition: an excited atomic state and a thermal field, varied across different average photon number values. Our results highlight the significant influence of the external classical field on the broadening of atomic lineshapes.

Furthermore, we have demonstrated that the external classical field can serve as a control signal for adjusting the width of the atomic lineshapes, even in the presence of a non-zero average temperature (see Fig.~\ref{Inversion atomica promedio 3d_alpha}). By precisely tuning the external classical driving field, it becomes feasible to finely manipulate the atomic transitions of the two-level system (qubit), which holds importance across various fields of research, including quantum information theory (for instance, see the final comment in Ref.~\cite{Brune_1996}), quantum communications (see the Section regarding Cavity QED in Ref.~\cite{Kimble_2008}), quantum optics (refer to Larson's exceptional book~\cite{Larson_2022}), spectroscopy, quantum control (see the description of Problem 7.3 in Nielsen's book~\cite{Nielsen_2010}), and quantum computing (refer, for instance, to Section 2.6.4 of Haroche and Raimond's book~\cite{Haroche_2006}).

Moreover, the consistency between analytical and numerical results underlines the robustness of our study. Additionally, we observe that these findings converge with those of the standard Jaynes-Cummings model when the interaction of the system with the classical field is disregarded. Ultimately, our research opens new perspectives for future investigations into the dynamics of quantum systems under the influence of external classical driving fields.

\section*{Acknowledgments}
L. Hernández-Sánchez thanks the Consejo Nacional de Humanidades, Ciencias y Tecnologías (CONAHCyT) by for the SNI III Assistant Scholarship (No. CVU: 736710). In turn, I. A. Bocanegra-Garay acknowledges the Spanish MCIN with funding from European Union Next Generation EU (PRTRC17.I1) and Consejeria de Educacion from Junta de Castilla y Leon through QCAYLE project, as well as grants PID2020-113406GB-I00 MTM funded by AEI/10.13039/501100011033, and RED2022-134301-T, also CONAHCyT (M\'exico) for financial support through project A1-S-24569, and to IPN (M\'exico) for supplementary economical support through the project SIP20232237.

%

\end{document}